\documentclass[sigconf]{acmart}

\AtBeginDocument{%
  \providecommand\BibTeX{{%
    \normalfont B\kern-0.5em{\scshape i\kern-0.25em b}\kern-0.8em\TeX}}}

\usepackage{multirow}
\usepackage{balance}
\usepackage[nolist]{acronym}
\begin{acronym}
\acro{MLM}{Masked Language Model}
\acro{NSP}{Next Sentence Prediction}
\end{acronym} 

\usepackage{tikz}
\usetikzlibrary{positioning,decorations.pathreplacing,calligraphy}

\copyrightyear{2022}
\acmYear{2022}
\setcopyright{acmlicensed}
\acmConference[NLBSE'22]{The 1st Intl. Workshop on Natural Language-based Software Engineering}{May 21, 2022}{Pittsburgh, PA, USA}
\acmBooktitle{The 1st Intl. Workshop on Natural Language-based Software Engineering (NLBSE'22), May 21, 2022, Pittsburgh, PA, USA}
\acmPrice{15.00}
\acmDOI{10.1145/3528588.3528661}
\acmISBN{978-1-4503-9343-0/22/05}

\begin{document}

\title{Predicting Issue Types with seBERT}


\author{Alexander Trautsch}
\email{alexander.trautsch@cs.uni-goettingen.de}
\affiliation{%
  \institution{University of Goettingen}
  \city{G\"{o}ttingen}
  \country{Germany}
}

\author{Steffen Herbold}
\email{steffen.herbold@tu-clausthal.de}
\orcid{0001-9765-2803}
\affiliation{%
  \institution{TU Clausthal}
  \city{Clausthal-Zellerfeld}
  \country{Germany}
}


\begin{abstract}
Pre-trained transformer models are the current state-of-the-art for natural language models processing. seBERT is such a model, that was developed based on the BERT architecture, but trained from scratch with software engineering data. We fine-tuned this model for the NLBSE challenge for the task of issue type prediction. Our model dominates the baseline fastText for all three issue types in both \textit{recall} and \textit{precision} to achieve an overall \textit{F1-score} of 85.7\%, which is an increase of 4.1\% over the baseline.
\end{abstract}



\maketitle

\section{Introduction}

Accurate issue type prediction can support developers and researchers~\cite{Herbold2020b}. Within this paper, we show how the pre-trained transformer model seBERT ~\cite{vondermosel2021validity} can be used for this purpose as part of the NLBSE 2022 tool competition~\cite{nlbse2022}. The seBERT model is a Natural Language Processing (NLP) model that was trained from scratch for the use within the software engineering domain, based on data from Stack Overflow, Jira and GitHub. seBERT uses the BERT$_\text{LARGE}$ architecture~\cite{devlin-etal-2019-bert}. In comparison, the original BERT was trained for the general domain and has sometimes problems with understanding words within the software engineering context~\cite{vondermosel2021validity}. We already showed on a different data set that seBERT outperforms BERT and fastText~\cite{fasttext} for the identification of bug issues~\cite{vondermosel2021validity}. We now transfer these results to the larger context of issue type prediction as a multi-class problem for the differentiation between the classes \textit{bug}, \textit{enhancement}, and \textit{question}.

\section{Issue Type Prediction with seBERT}

We now describe the data we used, our model architecture and training, the performance criteria we use, and the results we achieve. Our code for the experiments is available online.\footnote{https://github.com/atrautsch/nlbse2022\_replication\_kit}
We also provide a live version of the model.\footnote{https://user.informatik.uni-goettingen.de/~trautsch2/nlbse2022/}

\subsection{Data and Preprocessing}

The data we use consists of 722,899 issues collected from GitHub as development data, and 80,518 issues for testing~\cite{nlbse2022, ticket-tagger,  ticket-tagger-scp}. We have the following information for each issue:\footnote{The data also contains the URL of the issue and repository, the time of creation, and the author, but we do not use these fields.}

\begin{itemize}
  \item The label, which is one of the three classes we want to predict, i.e., \textit{bug}, \textit{enhancement}, and \textit{question}.
  \item The title of the issue.
  \item The description of the issue, also referred to as body.
\end{itemize}

We concatenate the title and body of the issue. Then, we replace line breaks with whitespaces and remove multiple subsequent whitespaces. Moreover, we encode the label as an integer, such that \textit{bug} is zero, \textit{enhancement} is one, and \textit{question} is two. Moreover, we split the development data into 80\% training data and 20\% validation data. 

\begin{table}
\caption{Overview of the amount of data for each class in the training and test data.}
\label{tbl:data}
\centering
\begin{tabular}{lrr}
\textbf{Class} & \textbf{\#Development} & \textbf{\#Test} \\
\hline\hline
\textit{bug} & 361,238 & 40,152 \\
\textit{enhancement} & 299,287 & 33,290 \\
\textit{question} & 62,373 & 7,076 \\
\hline
Total & 722,899 & 80,518 \\
\hline
\end{tabular}
\end{table}

\subsection{Model Architecture and Fine-tuning}

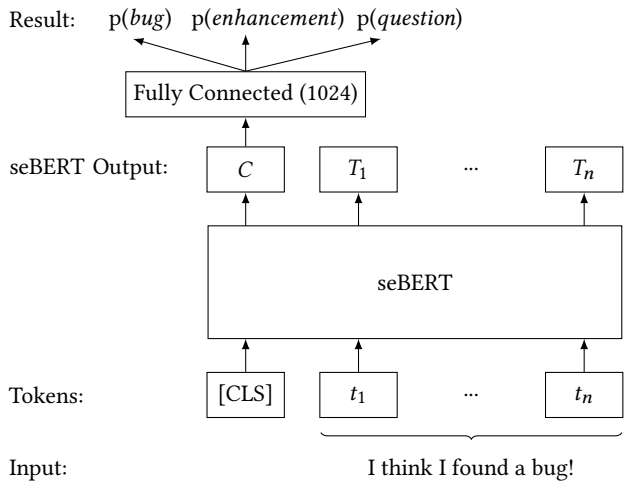
\begin{figure}
\begin{tikzpicture}
\node[shape=rectangle, minimum width={width("seBERT Output")+8pt}, text width={width("seBERT Output")+8pt}] (INPUT) at (0,-1) {Input:};
\node[shape=rectangle, minimum width={width("[CLS]")+8pt}, minimum height=0.6cm] (CLSIN) at ( 5,-1) {I think I found a bug!};
\draw [decorate, decoration = {brace, mirror}] (3,-0.5) --  (7,-0.5);

\node[shape=rectangle, minimum width={width("seBERT Output")+8pt}, text width={width("seBERT Output")+8pt}] (TOKENS) at (0,0) {Tokens:};
\node[shape=rectangle, draw=black, minimum width={width("[CLS]")+8pt}, minimum height=0.6cm] (CLSIN) at ( 2,0) {[CLS]};
\node[shape=rectangle, draw=black, minimum width={width("[CLS]")+8pt}, minimum height=0.6cm] (T1) at ( 3.5,0) {$t_1$};
\node[shape=rectangle, draw=black, minimum width={width("[CLS]")+8pt}, minimum height=0.6cm] (Tn) at ( 6.5,0) {$t_n$};
\node[shape=rectangle] (Tdots) at (5,0) {...};

\node [shape=rectangle, draw=black, minimum width=5.5cm, minimum height=1.5cm] (SEBERT) at (4.25,1.5) {seBERT};

\node[shape=rectangle, minimum width={width("seBERT Output")+8pt}, text width={width("seBERT Output")+8pt}] (OUTPUT) at (0,3) {seBERT Output:};
\node[shape=rectangle, draw=black, minimum width={width("[CLS]")+8pt}, minimum height=0.6cm] (COUT) at ( 2,3) {$C$};
\node[shape=rectangle, draw=black, minimum width={width("[CLS]")+8pt}, minimum height=0.6cm] (W1) at ( 3.5,3) {$T_1$};
\node[shape=rectangle, draw=black, minimum width={width("[CLS]")+8pt}, minimum height=0.6cm] (Wn) at ( 6.5,3) {$T_n$};
\node[shape=rectangle] (Wdots) at (5,3) {...};

\node[shape=rectangle, draw=black, minimum height=0.6cm] (FC) at (2, 4) {Fully Connected (1024)};

\node[shape=rectangle, minimum width={width("seBERT Output")+8pt}, text width={width("seBERT Output")+8pt}] (RESULT) at (0,5) {Result:};
\node[shape=rectangle, minimum width={width("p(\textit{enhancement})")+8pt}, text width={width("p(\textit{enhancement})")+8pt}] (BUG) at (0.5, 5) {p(\textit{bug})};
\node[shape=rectangle, minimum width={width("p(\textit{enhancement})")+8pt}, text width={width("p(\textit{enhancement})")+8pt}] (ENH) at (1.6, 5) {p(\textit{enhancement})};
\node[shape=rectangle, minimum width={width("p(\textit{enhancement})")+8pt}, text width={width("p(\textit{enhancement})")+8pt}] (QUE) at (3.8, 5) {p(\textit{question})};

\draw [-latex] (COUT.north) -- (FC.south);
\draw [-latex] (FC.north) -- (BUG.south);
\draw [-latex] (FC.north) -- (2, 4.8);
\draw [-latex] (FC.north) -- (QUE.south);

\draw [-latex] (CLSIN.north) -- (2, 0.75);
\draw [-latex] (T1.north) -- (3.5, 0.75);
\draw [-latex] (Tn.north) -- (6.5, 0.75);
\draw [-latex]  (2, 2.25) -- (COUT.south);
\draw [-latex]  (3.5, 2.25) -- (W1.south);
\draw [-latex]  (6.5, 2.25) -- (Wn.south);

\end{tikzpicture}
\caption{Overview of the architecture of our model.}
\label{fig:architecture}
\end{figure}

We use a transfer learning approach based on the fine-tuning of a pre-trained model. This means that we re-use a neural network that was trained for a different purpose. Figure~\ref{fig:architecture} shows an overview of our model which is based on the \texttt{BertForSequenceClassification} implementation of HuggingFace.\footnote{\url{https://huggingface.co/docs/transformers/model\_doc/bert\#transformers.BertForSequenceClassification}} 

We use seBERT~\cite{vondermosel2021validity} as foundation for our approach, a pre-trained BERT~\cite{devlin-etal-2019-bert} model which is based on the transformer architecture~\cite{Vaswani2017}. In comparison to the BERT model of Google, seBERT was trained from scratch based on data from the Software Engineering (SE) domain. Moreover, seBERT only has an input length of 128 tokens,\footnote{Words are tokenized based on the vocabulary. Words within the vocabulary are used as is, unknown words are represented by subword tokens.} in comparison to the 512 tokens from the original BERT models. Thus, our seBERT model only reasons about the first 128 tokens, i.e., only the title and the beginning of the issue text. BERT models use the [CLS] token and the output $C$ as placeholder for a classification task. $C$ is a 768  dimensional vector, that we use as input for one fully connected layer with 1024 neurons. The fully connected layer learns a logistic regression over $C$ and computes the probabilities of each class. We achieve this by using the cross-entropy loss~\cite{de2005tutorial}. For more details regarding seBERT and BERT we refer the to literature~\cite{devlin-etal-2019-bert, vondermosel2021validity}. 

We initialized our architecture with the weights of the pre-trained seBERT model. We then trained the model for five epochs on the training data. We did not freeze any weights, i.e., we did not just learn the weights of the fully connected layer, but also allowed updating the internal weights of seBERT. As optimizer, we used AdamW~\cite{loshchilov2018decoupled} with a learning rate of $lr=5\cdot 10^{-5}$, $\beta_1=0.9$ for the moving average of the gradients, $\beta_2=0.999$ for the moving average of the squared gradients, and no weight decay. We validated each epoch on the validation set and use the model with the highest validation accuracy, which was already achieved after the first epoch. No further hyperparameter tuning was applied.

\subsection{Performance Criteria}

The performance criteria we use are based on $tp_c, fn_c$, and $fp_c$, which are defined as follows. 

\begin{itemize}
    \item $tp_c$ are the true positive predictions of the class $c \in C=\{\textit{bug}, \textit{enhancement}, \textit{question}\}$. This means that the class $c$ is predicted correctly.
    \item $fn_c$ are the false negative predictions of the class $c$. This means that the class $c$ is missed and is instead predicted as a different class.
    \item $fp_c$ are the false positive predictions of class $c$. This means that a different class is wrongly predicted as class $c$. 
\end{itemize}

Based on these values, we define our performance criteria for each class $\textit{recall}_c, \textit{precision}_c$, and $\textit{F-Score}_c$:

\begin{equation}
\textit{recall}_c = \frac{tp_c}{tp_c+fn_c}
\end{equation}

\begin{equation}
\textit{precision}_c = \frac{tp_c}{tp_c+fp_c}
\end{equation}

\begin{equation}
\textit{F-Score}_c = 2\cdot\frac{\textit{recall}_c \cdot \textit{precision}_c}{\textit{recall}_c+\textit{precision}_c}
\end{equation}

The above criteria measure the performance for the individual classes. We use the \textit{micro average}\footnote{Please note that we use the micro average of the \textit{recall} directly, since the micro averages of \textit{recall}, \textit{precision}, and \textit{F-score} are equal. This is because $\sum_{c \in C} fp_c = \sum_{c \in C} fn_c$, which leads to the equality of the micro average of \textit{recall} and \textit{precision}. Since the \textit{F-Score} is their harmonic mean, it follows that is is also equal.} of these scores for our overall comparison of the performance:
\begin{equation}
    \textit{micro average} = \frac{\sum_{c \in C} tp_c}{\sum{c \in C} tp_c + fn_c}
\end{equation}

\subsection{Results}

Table~\ref{tbl:results} reports the performance of seBERT on the test data in comparison to the baseline model fastText. Overall, we achieve a \textit{micro average} score of 85.7\%, an improvement of 4.1\% over fastText. The results show that seBERT is superior to the baseline in all criteria, i.e., we are both better at correctly identifying issues of each class (\textit{recall}), as well as reducing the noise among these predictions (\textit{precision}). This means that the improvement is not at the cost of predictive power for one class, but benefits all classes. The largest improvement of seBERT is for the class \textit{question}, with an increase of 12.8\% in \textit{F-score}. However, we note that seBERT still misses over 50\% of the questions, i.e., there is still room for improvement. For the classes \textit{bug} and \textit{enhancement}, the performance of seBERT is much stronger, with \textit{F-scores} of 88.6\%, respectively 87.1\%. However, the difference to fastText is smaller, i.e., the improvement is only by 5.5\% for \textit{bug}, respectively 4.0\% for \textit{enhancement}. 

\begin{table}
\caption{Performance of seBERT in comparison to the baseline fastText.}
\label{tbl:results}
\begin{tabular}{l l c c c}
& & \textbf{fastText} & \textbf{seBERT} & \textbf{Difference} \\
\hline\hline
\multirow{3}{*}{\rotatebox[origin=c]{90}{\textit{recall}}}
& \textit{bug} & 81.6\% & 90.6\% & +9.0\% \\
& \textit{enhancement} & 84.5\% & 87.7\% & +3.2\% \\
& \textit{question} & 35.0\% & 48.7\% & +13.7\% \\
\hline
\multirow{3}{*}{\rotatebox[origin=c]{90}{\textit{precision}}}
& \textit{bug} & 83.1\% & 86.6\% & +3.5\% \\
& \textit{enhancement} & 81.6\% & 86.4\% & +4.8\% \\
& \textit{question} & 65.2\% & 73.1\% & +7.9\% \\
\hline
\multirow{3}{*}{\rotatebox[origin=c]{90}{\textit{F-score}}}
& \textit{bug} & 83.1\% & 88.6\% & +5.5\% \\
& \textit{enhancement} & 83.1\% & 87.1\% & +4.0\% \\
& \textit{question} & 45.6\% & 58.4\% & +12.8\% \\
\hline\hline
& \textit{micro average} & 81.6\% & 85.7\% & +4.1\% \\
\hline
\end{tabular}
\end{table}

\section{Threats to validity}

There is one notable threat to the internal validity of our results, due to the way seBERT was created. The corpus used for pre-training included the Github Issues (including comments) from 2015 to 2019~\cite{vondermosel2021validity}. The pre-training was self-supervised based on Masked-Language Modelling (MLM) and Next Sentence Prediction (NSP), i.e., did not include any information about the labels of the issues. However, we cannot rule out the possibility that the issue type prediction works better with data seen during the pre-training, because it is more likely that the language structure is already known by the model. Still, we believe that this threat is low: having seen the data before also increases the risk of memorization instead of generalization, which would actually decrease the performance on the test data.

\section{Conclusion}

We demonstrated the strong performance of the seBERT model for the task of issue type prediction over smaller models like fastText that were not pre-trained on a large corpus of data. While the performance is still not perfect, especially with respect to the detection of questions, results from earlier work demonstrate that humans are also not perfect at this task and there is likely some label noise~\cite{Herzig2013, Herbold2020b}. Still, there is room for improvement with pre-trained models. seBERT uses only a short input size of 128 tokens, which should be sufficient for most issues~\cite{vondermosel2021validity}, but may lead to missing vital information later in the issue for longer issues. Thus, models that are able to consider longer inputs may yield further improvements. For example, the recent Big Bird model is able to consider inputs with up to 4096 tokens~\cite{NEURIPS2020_c8512d14}. Moreover, we did not conduct extensive hyperparameter tuning to find the best possible training configuration (e.g., best learning rate) for seBERT, which means we possibly underestimate the capabilities of seBERT. 

\balance

\bibliographystyle{ACM-Reference-Format}
\bibliography{literature}

\end{document}